\documentclass[preprint,aps,nofootinbib]{revtex4}

\newcommand {\bc}{\begin{center}}
\newcommand {\ec}{\end{center}}
\newcommand {\bea}{\begin{eqnarray}}
\newcommand {\eea}{\end{eqnarray}}
\newcommand {\be}{\begin{equation}}
\newcommand {\ee}{\end{equation}}

\def\lsim{\mathrel{\rlap{\lower4pt\hbox{$\sim$}}
    \raise1pt\hbox{$<$}}}               
\def\gsim{\mathrel{\rlap{\lower4pt\hbox{$\sim$}}
    \raise1pt\hbox{$>$}}}  
\usepackage{graphicx}

\begin{document}


\title{Hydrodynamic tails and a fluctuation bound on 
the bulk viscosity}

\author{Mauricio Martinez and Thomas~Sch\"afer}

\affiliation{Department of Physics, North Carolina State University,
Raleigh, NC 27695}

\begin{abstract}
We study the small frequency behavior of the bulk viscosity spectral 
function using stochastic fluid dynamics. We obtain a number of model  
independent results, including the long-time tail of the bulk stress 
correlation function, and the leading non-analyticity of the spectral 
function at small frequency. We also establish a lower bound on the bulk 
viscosity which is weakly dependent on assumptions regarding the range 
of applicability of fluid dynamics. The bound on the bulk viscosity 
$\zeta$ scales as $\zeta_{\it min} \sim (P-\frac{2}{3}{\cal E})^2 \sum_i 
D_i^{-2}$, where $D_i$ are the diffusion constants for energy and 
momentum, and $P-\frac{2}{3}{\cal E}$, where $P$ is the pressure
and ${\cal E}$ is the energy density, is a measure of scale breaking.
Applied to the cold Fermi gas near unitarity, $|\lambda/a_s|\gsim 1$
where $\lambda$ is the thermal de Broglie wave length and $a_s$ is 
the $s$-wave scattering length, this bound implies that the ratio 
of bulk viscosity to entropy density satisfies $\zeta/s
\gsim 0.1\hbar/k_B$. Here, $\hbar$ is Planck's constant and 
$k_B$ is Boltzmann's constant.  
\end{abstract}

\maketitle

\section{Introduction}
\label{sec_intro}

 Hydrodynamic tails reflect the fact that fluid dynamics is an effective 
theory, in which the classical equations of motions are the lowest order 
approximation to a more complete theory involving averages over fluctuations 
of the fundamental variables. The classical equations of motion in fluid 
dynamics describe the evolution of conserved quantities such as mass, 
energy, and momentum. These equations depend on the form of the associated 
currents \cite{Kadanoff:1963}. In fluid dynamics the currents are expanded 
in gradients of hydrodynamic variables, and the corresponding expansion 
coefficients are known as transport coefficients. Transport coefficients 
control dissipative effects and fluctuation-dissipation relations imply 
that dissipative terms must be accompanied by stochastic forces. The 
presence of stochastic terms manifests itself in the form of long time, 
non-analytic, tails in correlation functions 
\cite{Dorfman:1965,Ernst:1971,Pomeau:1975}.

 Long time tails have been observed in computer simulations of 
fluids \cite{Alder:1967,Kadanoff:1989}, but they are more difficult to 
detect experimentally. In the present work we will study the correlation
function of the bulk stress, with an emphasis on dilute quantum 
fluids, such as the dilute Fermi gas near unitarity. Bulk stresses are 
interesting because the bulk viscosity can be strongly enhanced
near a phase transition \cite{Kadanoff:1968}, and quantum fluids
provide attractive applications because hydrodynamic fluctuations
are enhanced in systems in which the microscopic transport coefficients
are small. The existing literature contains only very limited information 
on the bulk stress correlation function. The only calculation of the bulk 
tail in a non-relativistic theory away from the critical point that we 
have been able to find appears to be wrong \cite{Pomeau:1975}. There are 
a number of studies of hydrodynamic tails near the liquid-gas endpoint 
and the superfluid transition \cite{Onuki:2002}, and there is a 
calculation of the bulk tail in a relativistic non-conformal fluid
at zero mean charge density in \cite{Kovtun:2003vj}.

 In this work we compute the long time tail of the bulk stress
correlation function in a non-relativistic fluid. We apply the 
result to the dilute Fermi gas near unitarity, and derive a novel
bound on the bulk viscosity of a non-conformal fluid. This bound
only depends on the shear viscosity and thermal conductivity of 
the fluid, combined with a measure of conformal symmetry breaking
in the equation of state. The bound is similar to lower bounds on 
the shear viscosity in relativistic and non-relativistic fluids that 
have been derived in \cite{Kovtun:2011np,Chafin:2012eq,Romatschke:2012sf}.
Finally, we discuss constraints on the bulk viscosity spectral function 
of a non-relativistic fluid. 

\section{Kubo Formula}
\label{sec_kubo}

 In this section we will determine the relation between the bulk 
viscosity and the low frequency behavior of the retarded correlation 
function of the stress tensor. This relation, known as the Kubo
formula, can be determined by matching the linear response relation
for the stress induced by an external strain to the low frequency 
behavior of the response predicted by fluid dynamics. The Kubo formula
for the shear and bulk viscosity of a non-relativistic fluid has been
rederived many times \cite{Landau:smII,Kadanoff:1963}, but there are 
a number of subtleties that we would like to emphasize. We will make
use of a formalism developed in 
\cite{Son:2005rv,Son:2005tj,Chao:2010tk,Chao:2011cy}, which is based 
on studying the response of the fluid to a non-trivial background metric 
$g_{ij}(t,\vec{x})$. Correlation functions of the stress tensor are 
determined using linear response theory, and the constraints of Galilean 
symmetry can be incorporated by requiring the equations of fluid dynamics 
to satisfy diffeomorphism invariance. 

 The retarded correlation function of the stress tensor $\Pi^{ij}$ is 
defined by
\be
\label{G_R}
G_R^{ijkl}(\omega,{\bf k}) = 
     -i \int dt \int d{\bf x}\, e^{i\omega t - i{\bf k \cdot x}} 
   \Theta(t) \langle [\Pi^{ij}(t,{\bf x}), \Pi^{kl}(0,{\bf 0})]\rangle \, . 
\ee
The retarded correlator determines the stress induced by a small
strain $g_{ij}(t,{\bf x})=\delta_{ij}+h_{ij}(t,{\bf x})$
\be
 \delta\Pi_{ij}(\omega,{\bf k}) = -\frac{1}{2}
 G_R^{ijkl}(\omega,{\bf k}) h_{kl}(\omega,{\bf k}).
\ee
In fluid dynamics the stress tensor is expanded in terms of gradients
of the thermodynamic variables. We write $\Pi_{ij} = \Pi_{ij}^0+\Pi_{ij}^1
+\ldots$, where
\be 
\Pi_{ij}^0 = \rho v_i v_j +P g_{ij} 
\ee
is the ideal fluid part, and $\Pi^i_{ij}$ with $i\neq 0$ are viscous 
corrections. Here, $\rho$ is the mass density of the fluid, $v_i$ is 
the velocity, and $P$ is the pressure. At first order in the gradient 
expansion $\Pi^{1}_{ij}=-\eta\sigma_{ij}-\zeta g_{ij}\langle\sigma\rangle$ 
with
\bea
\label{def_sig}
 \sigma_{ij} &=& \nabla_{i}v_{j}+\nabla_{j}v_{i}+\dot{g}_{ij}
         -\frac{2}{3}g_{ij}\langle\sigma\rangle \, , \\
\label{def_s}
 \langle\sigma\rangle &=& \nabla\cdot v+\frac{\dot{g}}{2g}\, , 
\eea
where $\sigma_{ij}$ is the shear stress tensor, $\eta$ is the shear 
viscosity, $\zeta$ is the bulk viscosity, $g$ is the determinant of 
the metric, and $\nabla_i$ is the covariant derivative associated with 
$g_{ij}$. The terms involving time derivatives of the metric are 
fixed by diffeomorphism invariance \cite{Son:2005tj}. Roughly, we 
can think of these terms as arising from the non-relativistic 
reduction of a generally covariant stress tensor, $\sigma_{ij}\sim
\nabla_i u_j\sim u_0\Gamma^0_{ij} \sim \dot{g}_{ij}$, where $u_0,u_i$ 
are the temporal and spatial components of the four-velocity, and 
$\Gamma^\alpha_{\mu\nu}$ is the Christoffel symbol. 

We will consider a harmonic perturbation of the form $h_{ij}(t,{\bf x}) = 
\delta_{ij}he^{-i\omega t}$. At the level of ideal fluid dynamics this
perturbation induces two terms in the stress tensor. The first, $\delta
\Pi_{ij}=Ph_{ij}$, arises from the direct coupling of $\Pi_{ij}^0$ to 
the background metric. The second term follows from the equations of
ideal fluid dynamics in a non-trivial background. The continuity
equation implies $\delta\rho = \frac{i\omega}{2}\, h\rho_0$, where 
$\rho_0$ is the unperturbed mass density. This leads to a shift in 
the pressure $\delta P= (\partial P)/(\partial \rho)_s\delta\rho$.

 At first order in gradients the response is carried by the coupling
to the background metric in equ.~(\ref{def_sig}). As expected, the 
response to a bulk strain $h_{ij}\sim \delta_{ij}$ is independent 
of the shear viscosity. At order $O(\omega)$ we get 
\be 
\label{Kubo_1}
 \frac{1}{9} G_R^{iijj}(\omega,{\bf 0}) = 
    -\left( \frac{2}{3}P - \left(\frac{\partial P}{\partial \rho}\right)_s
       \rho \right) - i\omega\zeta \, , 
\ee
where repeated indices are summed over. The Kubo relation is 
\be 
\label{Kubo_zeta}
 \zeta = -\lim_{\omega\to 0} \frac{1}{9\omega}\,{\rm Im} \, 
   G_R^{iijj}(\omega,{\bf 0 }) .
\ee
In the following we will derive a slightly more convenient version
of this Kubo relation. Bulk viscosity is a measure of scale breaking,
and we would like to find a version of the Kubo relation in which this
property is manifest. In the local rest frame of the fluid the trace
of the stress tensor is proportional to the pressure. In \cite{Chao:2011cy} 
we showed that in equilibrium scale breaking can be characterized by the 
quantity\footnote{We use the subscript ${\it Tr}$ to distinguish the trace 
anomaly $\Delta_{\it Tr} P$ from the quantity $\Delta P$, which is a 
fluctuation in the pressure.}
\be 
 \Delta_{\it Tr} P = P-\frac{2}{3}{\cal E}^0 \, .
\ee
Here, we use ${\cal E}^0$ to denote the energy density in the rest 
frame of the fluid. In ideal fluid dynamics the total energy density 
is given by ${\cal E}={\cal E}^0+\frac{1}{2}\rho {\bf v}^2$. We can
now make use of the fact that the energy density of the fluid is 
conserved
\be 
 \frac{\partial {\cal E}}{\partial t} 
    + {\boldmath \nabla}\cdot {\boldmath \jmath}^{\epsilon} = 0\, ,
\ee
where ${\boldmath \jmath}^\epsilon$ is the energy current. This 
relation implies that for $\omega\neq 0$ the retarded Green function 
$G_R^{\epsilon ii}(\omega,{\bf k})$ of the energy density and the trace 
of the stress tensor must vanish as ${\bf k}\to 0$. A more formal proof 
of this statement using Ward identities was given in \cite{Czajka:2017bod},
see also \cite{Jeon:1994if,Moore:2008ws}. We conclude that we can use any 
linear combination of the form  ${\cal O}= \frac{1}{3}\Pi^{ii}+c{\cal E}$ 
to define the Kubo relation for the bulk viscosity. Here, we will use 
$c=-\frac{2}{3}$. This choice has the nice property that the Kubo relation 
\be 
\label{Kubo_zeta_2}
 \zeta = -\lim_{\omega\to 0} \frac{1}{\omega}\,{\rm Im} \, 
   G_R^{{\cal O}{\cal O}}(\omega,{\bf 0 }) ,
   \hspace{1cm}
 {\cal O} =  \frac{1}{3}\left(\Pi^{ii}-2{\cal E}\right)
\ee
involves an operator which is manifestly sensitive to the 
trace anomaly in the hydrodynamic limit, ${\cal O} = \Delta_{\it Tr} P = 
P-\frac{2}{3}{\cal E}^0$.

\section{Hydrodynamic Fluctuations}
\label{sec_fluc}

 There are many possible strategies for evaluating the retarded correlation 
function of ${\cal O}=\Delta_{\it Tr} P$. An example is the microscopic
calculation in \cite{Schaefer:2013oba}, where we compute the bulk
viscosity in a dilute Fermi gas based on a perturbative calculation
of quasi-particle properties. In this work we will employ a different
strategy and compute the retarded correlation using a macroscopic 
theory of the long distance properties of the fluid. This theory 
it stochastic fluid dynamics \cite{Landau:smII}. As we will show this 
theory provides a universal prediction of the leading non-analyticity
in $G_R^{{\cal O}{\cal O}}(\omega,{\bf 0 })$ as $\omega\to 0$. It also 
provides a lower bound on $\zeta$, but this bound is sensitive to 
microscopic physics. 
 
 In order to explore the role of hydrodynamic fluctuations we 
will expand $\Delta P$ to second order in hydrodynamic variables. 
Higher order terms can be computed, but they provide corrections
that are subleading in $\omega/\omega_{\it br}$. Here $\omega_{\it br}$
is the breakdown scale of hydrodynamics, which we will define more
carefully below. The probability of a fluctuation of the hydrodynamic
variable is proportional to $\exp(\Delta S)$, where $\Delta S$ is 
the change in entropy of the fluid \cite{Landau:smI}. We can write 
\be 
  S = \int d^3{\bf x}\, s(\rho,{\cal E}^0) \, , 
\ee 
so that 
\bea
\Delta S &=& \int d^3{\bf x}\,\left\{ 
     \left(\frac{\partial s}{\partial \rho}\right)_{\!{\cal E}^0} \Delta\rho 
   + \left(\frac{\partial s}{\partial {\cal E}^0}\right)_{\!\rho} 
      \Delta{\cal E}^0 
   + \frac{1}{2}\left(\frac{\partial^2 s}{\partial\rho^2}\right)_{\!{\cal E}^0} 
           (\Delta\rho)^2 \right.
   \nonumber \\
  & & \hspace{2.00cm}\mbox{}\left.
   + \frac{\partial^2 s}{\partial \rho \partial {\cal E}^0}
         \Delta\rho\Delta {\cal E}^0
   + \frac{1}{2}\left(\frac{\partial^2 s}{\partial({\cal E}^0)^2}\right)_{\!\rho}
         (\Delta {\cal E}^0)^2
   + \ldots     \right\},
\eea  
We can use the conservation laws for the mass density $\rho$ and
the energy density ${\cal E}$ to show that the linear terms vanish. 
The quadratic terms can be simplified by using a set of thermodynamic
variables that diagonalizes the quadratic form. A suitable set of 
variables if provided by $(\rho,T)$ \cite{Landau:smII,Rubi:2000}. 
The entropy functional that governs fluctuations in $\rho,T$ and
${\bf v}$ is 
\be 
\label{s_fct}
\Delta S = -\frac{1}{2T_0}\int d^3{\bf x}\,\left\{ 
  \frac{1}{\rho_0}\left(\frac{\partial P}{\partial \rho}\right)_T
  (\Delta\rho)^2 + \frac{c_V}{T_0} (\Delta T)^2 
   + \rho_0 {\bf v}^2 + \ldots \right\}\, , 
\ee
where $(T_0,\rho_0)$ denote the mean values of the temperature and 
density, and $(\Delta T,\Delta \rho,{\bf v})$ are local fluctuations.
We can expand ${\cal O}=\Delta_{\it Tr} P$ to second order in 
$(\Delta T,\Delta\rho)$, 
\be 
\label{o_exp}
 {\cal O} = {\cal O}_0 + a_{\rho} \Delta \rho + a_T \Delta T
 + a_{\rho\rho}(\Delta\rho)^2 + a_{\rho T}\Delta\rho\Delta T
 + a_{TT}(\Delta T)^2 + \ldots \, . 
\ee
The hydrodynamic tails are determined by the second order terms. 
The corresponding coefficients can be expressed in terms of 
thermodynamic quantities. We find
\bea
 a_{\rho\rho} &=&  \frac{1}{2}\frac{\partial}{\partial\rho}
   \left[ c_T^2
      -\frac{2}{3}\left( \frac{h}{m} - \frac{T\alpha\kappa_T}{\rho} \right)
   \right]_T
\label{a_RR}     \, ,  \\
 a_{\rho T}   &=&  \left.\frac{\partial c_T^2}{\partial T}\right|_\rho
    -\frac{2}{3}  \left.\frac{\partial c_V}{\partial\rho}\right|_T
\label{a_RT} \, ,  \\
 a_{TT}      &=&  \frac{1}{2} \left[ 
         \frac{1}{T} \left(1-\rho\frac{\partial}{\partial\rho}\right)_T
    -\frac{2}{3}\left.\frac{\partial}{\partial T}\right|_\rho
      \right] c_V\, . 
\label{a_TT}
\eea
Here, $c_T$ is the isothermal speed of sound, $h$ is the enthalpy
per particle, $\alpha$ is the thermal expansion coefficient, $\kappa_T$
is the bulk modulus, and $c_V$ is the specific heat at constant 
volume. We define these quantities in the appendix. The coefficients
$a_{\alpha\beta}$ with $\alpha,\beta=(\rho,T)$ are sensitive to conformal 
symmetry breaking, and vanish in the ideal gas limit. A numerical estimate 
of $a_{\alpha\beta}$ therefore requires a non-trivial equation of state. As 
an example we consider a dilute Fermi gas governed by an $s$-wave
interacting with scattering length $a_s$. In the high temperature
limit the trace anomaly is given by \cite{Schaefer:2013oba}
\be
\label{Del_P_pert}
 \Delta_{\it Tr} P = \frac{2\pi}{3m^4a_s}\frac{\rho^2}{T}\, ,
\ee
where we employ units $\hbar=k_B=1$. 
In the limit $a_s\to\infty$ the dilute Fermi gas is scale invariant
and the trace anomaly vanishes. Using equ.~(\ref{Del_P_pert}) we find
\be 
 \left(a_{\rho\rho},a_{\rho T},a_{TT}\right) = 
   \frac{2\pi}{3m^4T^3a_s} \left( 
   T^2,-2\rho T, \rho^2\right)\, .  
\ee

\section{Hydrodynamic Tails: Formalism}
\label{sec_cor}

  In order to study hydrodynamic tails we consider the correlation
function of $\Delta_{\it Tr}P$ expanded to second order in $(\Delta\rho,
\Delta T)$. In statistical field theory it is convenient to start
from the symmetrized correlation function
\be 
 G_S^{{\cal OO}}(\omega,{\bf k}) = 
 \int d^3x\int dt\, e^{i(\omega t -{\bf k}\cdot{\bf x})} 
  \left\langle \frac{1}{2}
  \left\{ {\cal O}(t,{\bf x}), {\cal O}(0,0)\right\}
  \right\rangle \, . 
\ee
This function is related to the retarded correlator by the 
fluctuation-dissipation theorem. For $\omega\to 0$ we have 
\be 
\label{f-dis}
 G_S(\omega,{\bf k}) \simeq -\frac{2T}{\omega} 
  {\rm Im}\, G_R(\omega,{\bf k})\, . 
\ee
At second order in $(\Delta\rho,\Delta T)$ and at the level of the 
Gaussian entropy functional the symmetrized correlation function
factorizes into a set of two-point functions
\bea
\label{G_S_1l}
 G_S^{{\cal OO}}(\omega,0) &=& \int \frac{d\omega'}{2\pi}
   \int \frac{d^3{\bf k}}{(2\pi)^3} 
   \Big[ 2a_{\rho\rho}^2\Delta_S^{\rho\rho}(\omega',{\bf k}) 
                     \Delta_S^{\rho\rho}(\omega-\omega',{\bf k}) \\
  & & \hspace{1cm}\mbox{}
       + a_{\rho T}^2\Delta_S^{\rho\rho}(\omega',{\bf k}) 
                    \Delta_S^{TT}(\omega-\omega',{\bf k})
       + 2a_{TT}^2\Delta_S^{TT}(\omega',{\bf k}) 
                    \Delta_S^{TT}(\omega-\omega',{\bf k})
   \Big] \, . \nonumber 
\eea
where $\Delta_S^{\rho\rho}$ is the symmetrized density correlation function
\be 
\label{del_s_rho}
 \Delta_S^{\rho\rho}(\omega,{\bf k}) = 
 \int d^3x\int dt\, e^{i(\omega t -{\bf k}\cdot{\bf x})} 
  \left\langle \frac{1}{2}
  \left\{ \rho(t,{\bf x}), \rho(0,0)\right\}
  \right\rangle \, , 
\ee
and $\Delta_S^{TT}$ is the temperature correlation function. Note 
that by working with $(\Delta T,\Delta\rho)$ we avoid off-diagonal
correlation functions such as $\Delta_S^{\rho T}$. Also note that 
in hydrodynamics the symmetrized functions $\Delta_S$ reduces
to the statistical correlation function. 

 The Kubo relation involves the retarded, not the symmetrized,
correlation function. We can reconstruct the retarded function using 
the fluctuation-dissipation relation (\ref{f-dis}). Consider the first 
term in equ.~(\ref{G_S_1l}). At low frequency the contribution to $G_R$ 
can be written as \cite{Kovtun:2011np,Chafin:2012eq}
\bea
 \left. G_R^{{\cal OO}}(\omega,0)\right|_{\rho\rho} &=& 
     2a_{\rho\rho}^2 \int \frac{d\omega'}{2\pi}
   \int \frac{d^3{\bf k}}{(2\pi)^3}
   \Big[ \Delta_R^{\rho\rho}(\omega',{\bf k}) 
         \Delta_S^{\rho\rho}(\omega-\omega',{\bf k})
    \nonumber \\
  &&\hspace{4cm}\mbox{}       
         + \Delta_S^{\rho\rho}(\omega',{\bf k}) 
           \Delta_R^{\rho\rho}(\omega-\omega',{\bf k})
   \Big] .
\label{G_R_1l}
\eea
This is an example of a more general relation that one can prove using 
hydrodynamic effective actions, which shows that the retarded correlation 
functions can be derived using a perturbative expansion based on a 
combination of retarded and symmetrized propagators \cite{Martin:1973zz,Ma:1976,Hohenberg:1977ym,DeDominicis:1977fw,Khalatnikov:1983ak,Onuki:2002,Kovtun:2003vj,Kovtun:2012rj}.

The two-point functions of the temperature and density in first
order dissipative hydrodynamics are well known \cite{Kadanoff:1963}.
The temperature correlation function is dominated by a diffusive 
heat wave. The symmetric and retarded correlation functions are
\bea
\Delta^{TT}_S(\omega,{\bf k}) &=&  \frac{2T^2}{c_P}
  \frac{D_T {\bf k}^2}{\omega^2+(D_T{\bf k}^2)^2}\, , 
\label{Del_T_S}\\
\Delta^{TT}_R(\omega,{\bf k}) &=&   \frac{T}{c_P}
  \frac{-D_T{\bf k}^2}{-i\omega + D_T{\bf k}^2} \, ,
\label{Del_T_R}
\eea
where $c_P$ is the specific heat at constant pressure, $D_T=\kappa/c_P$ is 
the thermal diffusion constant, and $\kappa$ is the thermal conductivity.
The two-point function of the density is more complicated, because the
density couples to both propagating sound modes and diffusive heat 
modes. The symmetric correlation function is \cite{Kadanoff:1963}
\bea 
\Delta_S^{\rho\rho}(\omega,{\bf k}) &= & 2\rho T 
 \Bigg\{ \frac{\Gamma k^4}
        {\left(\omega^2-c_s^2k^2\right)^2+\left(\Gamma\omega k^2\right)^2}
 +   \frac{\Delta c_P}{c_s^2}
          \frac{D_T{\bf k}^2}
               {\omega^2+\left(D_Tk^2\right)^2} 
    \nonumber \\
 & & \hspace{0.5cm}\mbox{}
 -   \frac{\Delta c_P}{c_s^2}
 \frac{(\omega^2-c_s^2k^2)D_T{\bf k}^2}
      {\left(\omega^2-c_s^2k^2\right)^2+\left(\Gamma\omega k^2\right)^2}
 \Bigg\} \, ,
\label{Del_S_L}
\eea
where $k^2={\bf k}^2$, $c_s$ is the speed of sound, and $\Delta c_P=(c_P-c_V)
/c_V$. We have also defined the sound attenuation constant
\be
\label{Gam}
 \Gamma = \frac{4}{3}\frac{\eta}{\rho}+\frac{\zeta}{\rho} 
   + \kappa\left(\frac{1}{c_V}-\frac{1}{c_P}\right) 
  = \frac{4}{3}\frac{\eta}{\rho}\left[
     1 + \frac{3}{4}\frac{\zeta}{\eta}
       + \frac{3}{4}\frac{\Delta c_P}{{\it Pr}} \right]\, ,
\ee
where ${\it Pr}=(c_P\eta)/(\rho\kappa)$ is the Prandtl number, the ratio of 
the momentum and thermal diffusion constants. At high temperature $\Delta c_P
=2/3$ and ${\it Pr}=2/3$ \cite{Braby:2010ec}, and at low temperature 
$\Delta c_P/{\it Pr}\to 0$. 

 The two-point function of the density has a complicated pole
structure, and the calculation of loop diagrams can be simplified
by separating the different terms. We will also separate the 
contributions from sound and diffusive modes, 
\be 
\label{Del_nn_sum}
 \Delta^{\rho\rho}_{R,S}(\omega,{\bf k}) 
   =  \Delta^{\it sd}_{R,S}(\omega,{\bf k})
    + \Delta^{\it ht}_{R,S}(\omega,{\bf k})
    + \Delta^{\it m}_{R,S}(\omega,{\bf k})\, . 
\ee
In the long wavelength limit the sound contribution can be 
written as 
\bea 
\Delta^{\it sd}_{S}(\omega,{\bf k}) &=& 
  \rho T\, \frac{\Gamma k^3}{2\omega c_s}
    \left\{ \frac{1}{(\omega-c_sk)^2+(\frac{\Gamma}{2} k^2)^2}
           -\frac{1}{(\omega+c_sk)^2+(\frac{\Gamma}{2} k^2)^2}
    \right\} 
\label{Del_r_sd_S}\\
\Delta^{\it sd}_{R}(\omega,{\bf k}) &=& 
    \;\;   \rho \, \frac{\Gamma k}{2c_s}\;
       \left\{\; \frac{1}{\omega-c_sk+i\frac{\Gamma}{2} k^2}\;
             -\; \frac{1}{\omega+c_sk+i\frac{\Gamma}{2} k^2}\;
    \right\}\, 
\label{Del_r_sd_R}
\eea
and the diffusive heat mode is 
\bea
\Delta^{\it ht}_{S}(\omega,{\bf k}) &=& 
  2\rho T\,  \frac{\Delta c_P}{c_s^2}
          \frac{D_T{\bf k}^2}
               {\omega^2+\left(D_Tk^2\right)^2} \, ,  
\label{Del_r_ht_S}\\
\Delta^{\it ht}_{R}(\omega,{\bf k}) &=& 
    \rho \,  \frac{\Delta c_P}{c_s^2}
          \frac{-D_T{\bf k}^2}
               {-i\omega + D_Tk^2} \, .
\label{Del_r_ht_R}
\eea
Finally, there is a term that is sensitive to both sound and
diffusive modes 
\bea
\Delta^{\it m}_{S}(\omega,{\bf k}) &=& 
 -2\rho T\,  \frac{\Delta c_P}{c_s^2}
          \frac{kD_T}{2c_s}
   \left\{ \frac{\omega-c_sk}{(\omega-c_sk)^2+(\frac{\Gamma}{2} k^2)^2}
          -\frac{\omega+c_sk}{(\omega+c_sk)^2+(\frac{\Gamma}{2} k^2)^2}
         \right\}\, ,  
\label{Del_r_m_S}\\
\Delta^{\it m}_{R}(\omega,{\bf k}) &=& 
   \; \rho \;  \frac{\Delta c_P}{c_s^2}
         \frac{i\omega kD_T}{2c_s}\,
   \left\{\; \frac{1}{\omega-c_sk+i\frac{\Gamma}{2} k^2}\;
             -\; \frac{1}{\omega+c_sk+i\frac{\Gamma}{2} k^2}\;
    \right\}\, .
\label{Del_r_m_R}
\eea

\begin{figure}[t]
\bc\includegraphics[width=11cm]{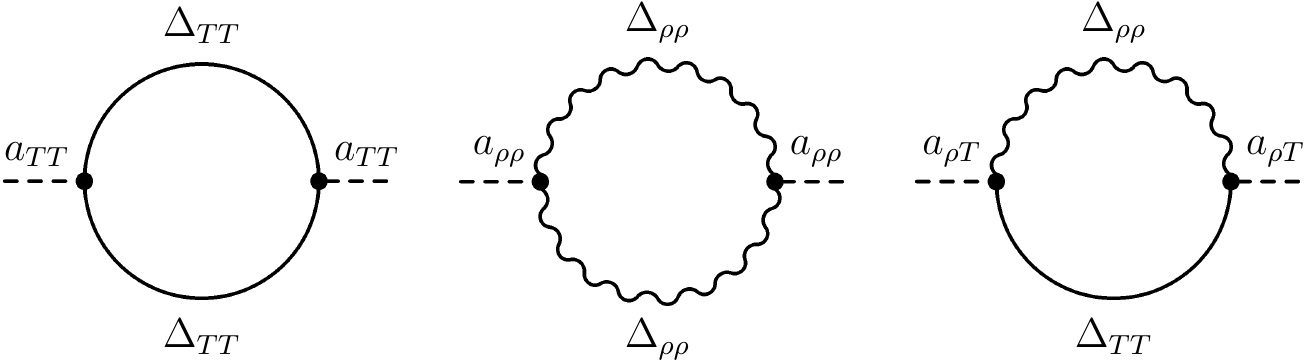}\ec
\vspace*{-4cm}\flushleft\hspace*{2cm} a)\hspace*{3.6cm}b)\hspace*{3.6cm}c)
\vspace*{3.0cm}
\caption{\label{fig_loop}
Diagrammatic representation of the leading contribution of thermal 
fluctuations to the bulk stress correlation function. The dashed line
corresponds to the operator ${\cal O} = P-\frac{2}{3}{\cal E}$. Solid 
lines denote the diffusive temperature correlator, and wavy lines 
denote the density correlation function, determined by the sound pole 
and the diffusive heat mode.}   
\end{figure}

\section{Hydrodynamic Tails: One-loop diagrams}
\label{sec_loops}

 In this section we will compute the leading infrared behavior of the 
three one-loop diagrams shown in Fig.~\ref{fig_loop}. The two-point 
function of the density has three distinct contributions,  see 
equ.~(\ref{Del_nn_sum}), and as a result there are ten one-loop
diagrams total. As we will see, only four of them contribute to the 
low frequency behavior of $G_R(\omega,{\bf 0})$. 

 1. The simplest diagram involves diffusive fluctuations of the 
temperature only. We consider equ.~(\ref{G_R_1l}) with $(\rho\rho)\to 
(TT)$ and use the retarded and symmetrized functions given in 
equ.~(\ref{Del_T_S},\ref{Del_T_R}). We perform the frequency integral
by closing the contour in the complex $\omega$ plane. We find
\be 
\label{G_R_1}
\left.G_R^{\cal OO}(\omega,{\bf 0})\right|_{TT}^{\it ht} 
  = - \frac{2a_{TT}^2T^3}{c_P^2} 
   \int \frac{d^3k}{(2\pi)^3} \, \frac{k^2}{k^2-\frac{i\omega}{2D_T}}\, , 
\ee
where $TT$ refers to the presence of two temperature correlation
functions, and ${\it ht}$ indicates that these modes are dominated 
by a diffusive heat mode. The integral in equ.~(\ref{G_R_1}) is 
ultraviolet divergent. We will regularize the 
integral using a momentum cutoff $\Lambda$. We will see that there 
are two types of terms. Hydrodynamic tails are non-analytic in 
$\omega$ and independent of the cutoff. Fluctuation terms are 
sensitive to the cutoff and contribute to $G_R$ in the same way
as transport coefficients. This implies that the cutoff dependence
can be absorbed into the bare transport parameters. However, we
will see that this procedure implies bounds on the transport
coefficients. 

 After introducing a cutoff we can compute the integral in 
equ.~(\ref{G_R_1}) by expanding in $\omega$. The leading terms are 
\be 
\label{G_R_2}
\left.G_R^{\cal OO}(\omega,{\bf 0})\right|_{TT}^{\it ht} 
  =- \frac{2a_{TT}^2T^3}{c_P^2}
     \, L(\omega,\Lambda,2D_T)\, , 
\ee
where we have defined 
\be   
\label{loop_fct}
 L(\omega,\Lambda,2D_T)= \frac{1}{2\pi^2}\left\{
   \frac{\Lambda^3}{3}+\frac{i\omega\Lambda}{2D_T} 
   -\frac{\pi}{2\sqrt{2}}(1+i) \left(\frac{\omega}{2D_T}\right)^{3/2}
   + \ldots \right\} \, . 
\ee 
Note that the small parameter in the low frequency expansion is 
$\epsilon\equiv \omega/(D_T\Lambda^2)$.
We observe that the $\Lambda^3$ term can be viewed as a contribution
to the compressibility term in equ.~(\ref{Kubo_1}), and the $i\omega
\Lambda$ term is a contribution to the bulk viscosity. This term
is sensitive to scale breaking via the coefficient $a_{TT}$, and
it scales inversely with the thermal conductivity. The last term
is a hydrodynamic tail. The imaginary part can be viewed as a
$\sqrt{\omega}$ contribution to the frequency dependent bulk 
viscosity $\zeta(\omega)$, and real part is a $1/\sqrt{\omega}$ 
contribution the bulk viscosity relaxation time. This term signals the 
breakdown of second order deterministic fluid dynamics in the 
low frequency limit. 

2. A similar diffusive heat contribution appears in the two point 
function of the density. Comparing equ.~(\ref{Del_T_S},\ref{Del_T_R})
to equ.~(\ref{Del_r_ht_S},\ref{Del_r_ht_R}) we observe that this 
contribution is equal to the previous term up to an overall factor.
We get 
\be 
\label{G_R_3}
\left.G_R^{\cal OO}(\omega,{\bf 0})\right|_{\rho\rho}^{\it ht} 
=- \frac{2a_{\rho\rho}^2T\rho^2(\Delta c_P)^2}
     {c_s^2}\,  L(\omega,\Lambda,2D_T)
\ee 
In the case of a dilute gas equ.~(\ref{G_R_3}) and equ.~(\ref{G_R_2})
are comparable in magnitude, but in general the two contributions
can be different.

3. Another diffusive heat contribution is contained in the mixed
$\Delta_{\rho\rho}\Delta_{TT}$ term, shown as the third diagram in 
Fig.~\ref{fig_loop}. We get 
\be 
\label{G_R_4}
\left.G_R^{\cal OO}(\omega,{\bf 0})\right|_{\rho T}^{\it ht} 
 =- \frac{a_{\rho T}^2\rho T^2 \Delta c_P}
{c_P c_s^2}\,  L(\omega,\Lambda,2D_T)\, . 
\ee 

4. The two point function of the density also contains a sound 
contribution. This term is quite different, because sound is a propagating
mode, and sound attenuation is controlled by $\Gamma$, which is not 
only sensitive to $\kappa$ but also to the shear viscosity $\eta$ 
and a possible microscopic contribution to $\zeta$. We determine
this term using the two point functions in equ.~(\ref{Del_r_sd_S},
\ref{Del_r_sd_R}). We observe that there are two types of contributions,
characterized by the relative sign of the real part of the pole 
position, $\omega_\pm^\prime=\pm c_s k + O(\omega,k^2)$. We first
consider diagrams where the poles are on opposite sides of the 
real axis. We get 
\be 
\label{G_R_5}
\left.G_R^{\cal OO}(\omega,{\bf 0})\right|_{\rho\rho}^{\it sd} 
  =- \frac{a^2_{\rho\rho}T\rho^2}
{c_s^4}  \, L(\omega,\Lambda,\Gamma)\, ,
\ee
where the index ${\it sd}$ indicates the contribution from the sound 
mode. The diagram where the two poles are on the same side gives
\be
\label{G_R_5b}
\left.G_R^{\cal OO}(\omega,{\bf 0})\right|_{\rho\rho}^{\it sd} 
 =- \frac{a^2_{\rho\rho}T\rho^2}{4c_s^2}
   \int \frac{d^3k}{(2\pi)^3} 
   \frac{k^2}{(\omega-2c_sk+i\Gamma k^2)(c_sk-i\frac{\Gamma k^2}{2})}\, . 
\ee
This integral is UV divergent, but it is less IR sensitive then 
equ.~(\ref{G_R_1}). In particular, the low frequency behavior is 
governed by $c_sk\gg \Gamma k^2$. As a result, the contribution 
to the $i\omega$ term in $G_R^{\cal OO}(\omega,{\bf 0})$ is suppressed
by a factor $(\Gamma\Lambda/c_s)$ relative to equ.~(\ref{G_R_2}).

5. The remaining diagrams fall into two categories. The first class involves 
mixed diagrams in which a diffusive heat mode is coupled to a propagating 
sound mode. These diagrams are suppressed because if one of the propagators 
is put on shell the other propagator is far off shell, and the diagram is not 
infrared sensitive. The other diagrams involve the mixed sound-heat propagator 
in equ.~(\ref{Del_r_m_S},\label{Del_r_m_R}). The on-shell residue of this
propagator is suppressed. We finally collect the contributions from 
equ.~(\ref{G_R_2}-\ref{G_R_6}). We get 
\be 
\label{G_R_6}
G_R^{\cal OO}(\omega,{\bf 0}) =-A_T L(\omega,\Lambda,2D_T)
  - A_\Gamma L(\omega,\Lambda,\Gamma)\, , 
\ee
where we have defined
\be
\label{A_TG} 
A_T =    \frac{2a_{TT}^2T^3}{c_P^2} 
         + \frac{2a_{\rho\rho}^2\rho^2 T(\Delta c_P)^2}{c_s^4}
         + \frac{a_{\rho T}^2\rho T^2\Delta c_P}{c_P c_s^2}\, , 
\hspace{0.75cm}
A_\Gamma =  \frac{a_{\rho\rho}^2\rho^2 T}{c_s^4}\, . 
\ee

\section{Phenomenological estimates}
\label{sec_num}
\subsection{Hydrodynamic tail}
\label{sec_tail}

 In the previous section we showed that the $\omega^{3/2}$ term
in the retarded correlation function is uniquely determined in 
terms of the equation of state and the transport parameters.
This term has several physical effects: It determines the long
time tail of the correlation function, it governs the small frequency
limit of the bulk viscosity spectral function, and it determines
the $\omega\to 0$ divergence in the relaxation time. We first
consider the correlation function
\be 
 C_{\zeta}(t) = \int \frac{d\omega}{2\pi}\, G_S^{{\cal OO}}(\omega,{\bf 0}) 
  e^{-i\omega t}\, . 
\ee
For $t\to\infty$ we obtain a $t^{-3/2}$ tail
\be 
 C_\zeta(t) = \frac{T}{4\pi^{3/2}}
\left( \frac{A_T}{(2D_T)^{3/2}} + \frac{A_\Gamma}{\Gamma^{3/2}}
     \right) \frac{1}{t^{3/2}}\, ,
\ee
This contribution is computed most easily by starting from the momentum
integral in equ.~(\ref{G_R_1}), and then perform the frequency integral
before the momentum integral. The hydrodynamic tail in the bulk stress 
correlator was first computed by Pomeau and R\'esibois \cite{Pomeau:1975}, 
but their result does not appear to be correct. In particular, the 
expression for $C_{\zeta}(t)$ given in \cite{Pomeau:1975} does not vanish 
for a scale invariant fluid. In our work $C_\zeta(t) \sim a_{\alpha\beta}^2 
\sim (\Delta_{\it Tr}P)^2$ automatically vanishes for a scale invariant fluid. 

 The contribution of critical fluctuations to the tail in the bulk
stress correlation function was computed by Onuki \cite{Onuki:2002}, 
both in model H (liquid-gas endpoint) and model F (superfluid transition). 
In principle the model F result for $T>T_c$ is directly applicable to the 
Fermi gas near unitarity. Model F contains two hydrodynamic variables, a 
linear combination of the energy density ${\cal E}$ and the density $\rho$, 
as well as the the superfluid density. Above $T_c$ only the energy density 
like variable contributes. In this regime there are two differences compared 
to our analysis: 1) We keep both both ${\cal E}$ and $\rho$; 2) The model 
F analysis uses a more complicated functional form of the thermal 
conductivity $\kappa(k^2,t)$ with $t=(T-T_c)/T_c$, which reduces to a 
simple constant for $t\gg 1$. This implies that the model F tail should
be similar to our tail for large $t$. This is difficult to verify, 
because the coupling between the energy density-like variable to the bulk 
stress does not manifestly respect scale invariance. The bulk tail in a 
relativistic non-conformal fluid was computed by Kovtun and Yaffe 
\cite{Kovtun:2003vj}. These authors assume that the mean density of 
the fluid vanishes, so that we cannot directly compare to the non-relativistic
limit.

\subsection{Spectral function}
\label{sec_spec}

 A second quantity of interest is the spectral function
\be 
 \zeta(\omega) =-\frac{1}{9\omega} {\rm Im}G_R^{{\cal OO}}(\omega,{\bf 0})\, . 
\ee
The existence of a hydrodynamic tail implies that 
\be 
\label{zeta_w_h}
 \zeta(\omega) = \zeta(0) - 
 \left( \frac{A_T}{(2D_T)^{3/2}} + \frac{A_\Gamma}{\Gamma^{3/2}}\right)
 \frac{\sqrt{\omega}}{36\sqrt{2}\pi} \, .
\ee 
This result can be combined with other model independent information
about the spectral function. The high frequency tail of the bulk viscosity 
was determined using the operator product expansion \cite{Hofmann:2011qs}
\be 
\label{zeta_w_t}
 \zeta(\omega) = \frac{{\cal C}}{36\pi\sqrt{m\omega}}
  \frac{1}{1+a_s^2m\omega}\, ,
\ee
where ${\cal C}$ is the contact density \cite{Tan:2005,Tan:2008}. The 
contact density is directly related to the trace anomaly near unitarity
\be 
\Delta_{\it Tr}P = \frac{{\cal C}}{12\pi ma_s}\, .
\ee
In the high temperature limit ${\cal C}$ can be computed using the 
virial expansion \cite{Yu:2009}. Near unitarity we find
\be
 {\cal C} = 4\pi n^2\lambda^2 \left\{ 1 + 
  \frac{1}{\sqrt{2}}\left(\frac{\lambda}{a_s}\right) + \ldots \right\}\, , 
\ee
which implies
\be 
\zeta(\omega) \sim \lambda^{-3}\left(\frac{z\lambda}{a_s}\right)^2
  \left( \frac{T}{\omega}\right)^{3/2}\, , 
\ee
where $\lambda=[(2\pi)/(mT)]^{1/2}$ is the thermal de Broglie wave length
and $z=\frac{1}{2}n\lambda^3$ is the fugacity of the gas. Finally, there 
is a sum rule for the bulk viscosity spectral function
\cite{Taylor:2010ju,Enss:2010qh,Goldberger:2011hh}
\be 
\label{sum_rule}
 \frac{1}{\pi} \int d\omega\, \zeta(\omega) = 
  \frac{1}{72\pi ma^2} 
  \left. \frac{\partial {\cal C}}{\partial a_s^{-1}}\right|_{s/n}\, .
\ee
In the next section we will combine these constraints with the 
fluctuation bound to provide a simple model of the viscosity
spectral function.

\begin{figure}[t]
\bc\includegraphics[width=11cm]{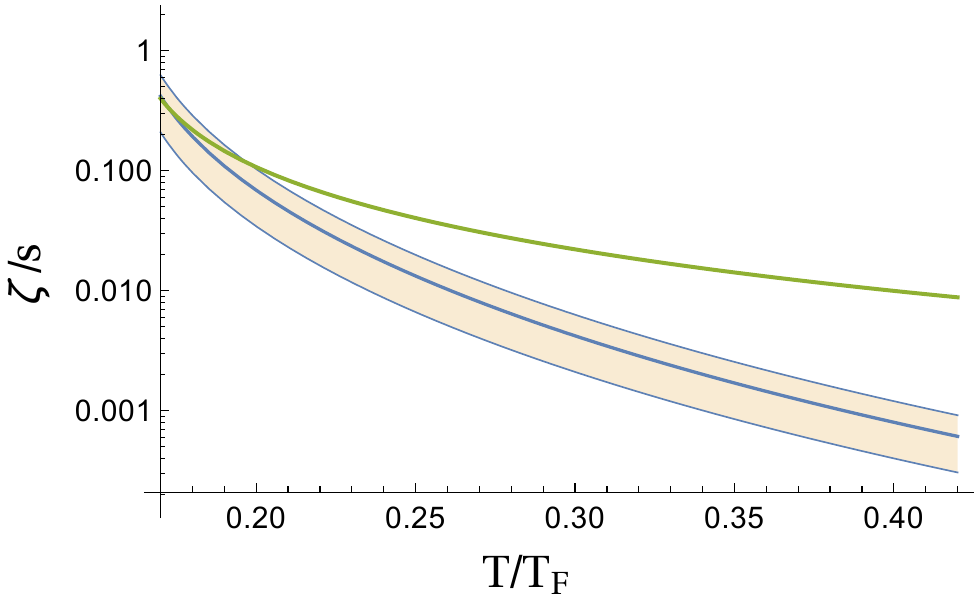}\ec
\caption{\label{fig_bound}
Fluctuation bound (blue line) on $\zeta/s$ for the dilute Fermi gas 
as a function of $T/T_F$. We show the regime $T>T_c$ with $T_c/T_F\simeq
0.17$. As explained in the text we estimate the equation of state and 
transport properties using results in the high temperature limit. We 
have also chosen $a_s/\lambda=1$. The error band corresponds to a $50\%$ 
error in $\Lambda_T$ and $\Lambda_\Gamma$. For comparison we show the 
kinetic theory result for $\zeta/s$ as the green line.}   
\end{figure}

\subsection{Fluctuation bound}
\label{sec_bound}

 The cutoff dependent term in the bulk viscosity 
\be 
 \zeta_\Lambda = \frac{1}{18\pi^2}\left( \frac{A_T\Lambda}{2D_T} 
   + \frac{A_\Gamma\Lambda}{\Gamma}
     \right)  \, ,
\ee
has to combine with the bare bulk viscosity to determine the physical
bulk viscosity of the fluid. We can view this result as arising from 
a renormalization group procedure, where fluid dynamics is matched to 
a microscopic theory at the scale $\Lambda$, and then the evolution 
of $G_R(\omega)$ below the scale $\Lambda$ is computed using stochastic
fluid dynamics. For this procedure to be consistent the bare viscosity
at the cutoff scale must be positive, and the the physical viscosity
must be larger than $\zeta_\Lambda$. This bound increases with the 
cutoff scale $\Lambda$. The largest possible $\Lambda$ is determined
by the breakdown scale of fluid dynamics, because above that scale
stochastic fluid dynamics is not reliable. Of course, the viscosity 
at the cutoff scale must depend on $\Lambda$, so that the physical 
viscosity $\zeta(0)$ is cutoff independent. The same conclusion also 
follows from the spectral density given in equ.~(\ref{zeta_w_h}). We
observe that the non-analytic $\sqrt{\omega}$ term is negative. If 
this term is the dominant correction to the spectral density below
the breakdown scale of fluid dynamics, $\omega\lsim \omega_{\it br}$, 
then spectral positivity implies that $\zeta(0)$ cannot be arbitrarily
small. 

 In order to determine the maximum momentum where fluid dynamics
can be trusted we can study the dispersion relation of diffusive
heat modes and propagating sound waves, and determine the maximum
momentum for which higher order corrections are small compared to
leading order terms. 

\begin{figure}[t]
\bc\includegraphics[width=11cm]{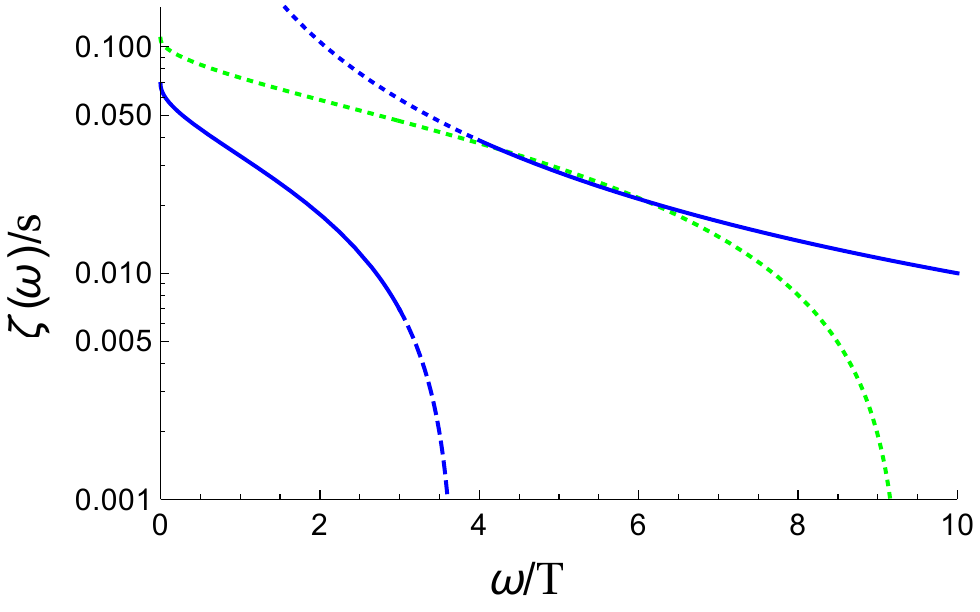}\ec
\caption{\label{fig_spec}
Schematic form of the bulk viscosity spectral function. This figure
shows $\zeta/s$ as a function of the frequency $\omega$ in units of 
$T$. We have chosen $a_s/\lambda=1$ and $T/T_F=0.2$. The low frequency
part shows the function $\zeta(\omega)=\zeta_{\it min}-c\sqrt{\omega}$,
where $\zeta_{\it min}$ is the bound in equ.~(\ref{zeta_bound}), and $c$ 
is the universal coefficient given in equ.~(\ref{zeta_w_h}). The high 
frequency part is the function given in equ.~(\ref{zeta_w_t}). The green
dotted line shows a model for the low frequency spectral function where
we have added a microscopic contribution $\zeta_{\it micro}/s =0.04$ to 
the hydrodynamic result. The microscopic contribution was chosen to 
smoothly match the high frequency tail.}  
\end{figure}

1. Diffusive modes: Heat modes are characterized by $\omega\sim 
D_T k^2$. Corrections arise from higher order terms in the derivative
expansion. For non-zero frequency the leading correction is due to the 
relaxation time. We get $\omega \sim D_T k^2 \ll \tau_\kappa^{-1}$. For 
this relation to be maintained for all $k<\Lambda$ we have to require 
that $\Lambda \lsim \Lambda_T$ with $\Lambda_T=(\tau_\kappa D_T)^{-1/2}$. 
In kinetic theory $\tau_\kappa=(m\kappa)/(c_PT)$ and 
\be 
\label{Lam_T}
 \Lambda_T \simeq \frac{1}{D_T}\left(\frac{T}{m}\right)^{1/2}\, . 
\ee
Equation (\ref{Lam_T}) implies that the expansion parameter of the low 
frequency expansion, $\epsilon=\omega/(D_T\Lambda^2)$, is of order $\epsilon 
\sim (mD_T)(\omega/T)$. For a nearly perfect fluid $D_T\sim m^{-1}$ 
\cite{Schafer:2009dj} and the low frequency expansion is valid all the way 
up to $\omega\sim T$. In the case of a poor fluid $D_T\gg m^{-1}$
and the range of validity of the low frequency expansion is smaller. 
We also note that equ.~(\ref{Lam_T}) ensures that the expansion 
parameter $(D_T\Lambda/c_s)$ is indeed small.
 
2. Sound channel: In the sound channel we have $\omega \sim c_s k \ll 
\Gamma k^2$. This implies $k\lsim \Lambda_\Gamma$ with
\be 
\label{Lam_sd}
 \Lambda_\Gamma \simeq \frac{1}{\Gamma}
    \left(\frac{\partial P}{\partial\rho}\right)_{s/n}^{1/2}\, . 
\ee
For a weakly interacting gas we get $(\partial P)/(\partial\rho)_{s/n}
\simeq (5T)/(3m)$. We can either use the two estimates equ.~(\ref{Lam_T},
\ref{Lam_sd}) in the respective channels, or use the smaller of the 
two values. In the weak coupling limit, where ${\it Pr}\sim 1$, these
two estimates are numerically very similar. Using the first method, we 
obtain the bound 
\be 
\label{zeta_bound}
 \zeta_{\it min} = \left( \frac{A_T}{2D_T^2} 
                      + \frac{\sqrt{5}A_\Gamma}{\sqrt{3}\Gamma^2}
     \right) \sqrt{\frac{T}{m}} \, .
\ee
We observe that there is a minimum value of $\zeta$ that is solely
controlled by $(\Delta_{\it Tr}P/D_T)^2$ and $(\Delta_{\it Tr}P/\Gamma)^2$.
This implies that if there is scale breaking in the equation of state, 
and if the shear viscosity and thermal conductivity are finite, then 
the bulk viscosity cannot be zero. Fluctuation bounds on the shear 
viscosity were studied in \cite{Chafin:2012eq,Romatschke:2012sf}. We
observe that the bound on $\zeta$ has the same structure as the 
bound on $\eta$, but is suppressed by a factor $(\Delta_{\it Tr}P/P)^2$.
 
 Finally, we provide some numerical estimates. For this purpose we 
assume that the bare bulk viscosity is zero, and that the shear viscosity 
and thermal conductivity are described by kinetic theory, $\eta=\eta_0
(mT)^{3/2}$ and $\kappa=\kappa_0m^{1/2}T^{3/2}$ with $\eta_0=15/(32\sqrt{\pi})$ 
and $\kappa_0=225/(128\sqrt{\pi})$ \cite{Bruun:2007b,Braby:2010ec}. In the 
case of the shear viscosity this is known to be a good approximation even 
close to the critical temperature \cite{Bluhm:2017rnf}. We also use the 
results for $c_s^2$, $c_P$ and $\Delta c_P$ in the dilute limit, see 
Appendix \ref{sec_app}. The bound on $\zeta/s$ as a function of $T/T_F$ 
is shown in Fig.~\ref{fig_bound}. The width of the band reflects a $50\%$ 
error related to the choice of $\Lambda$. For comparison we also show the 
kinetic theory result $\zeta/n=z^2/(24\sqrt{2}\pi\lambda^3)(\lambda/a_s)^{2}$ 
\cite{Schaefer:2013oba}. At high temperature the fluctuation bound is very 
small, but near $T_c\simeq 0.17T_F$ \cite{Ku:2011} the bound is comparable 
to the kinetic theory result, indicating that the bulk viscosity must be 
at least as big as predicted by kinetic theory. Note that we have 
extrapolated the bound on $\zeta/s$ all the way to $T_c$, despite the 
fact that several estimates involve approximations that are only reliable 
for $T\gg T_c$. Similar to the kinetic theory estimates discussed above, 
it is known that in the case of $\eta/s$ this procedure provides a 
numerically accurate estimate of the bound near $T_c$.

 We note that $\zeta/s$ is given in units of $\hbar/k_B$. Both the 
hydrodynamic and the kinetic theory calculation are completely 
classical. Planck's constant enters the hydrodynamic calculation
via the equation of state, and it appears in the kinetic theory
calculation in terms of both the equation of state and the 
quasi-particle dispersion relation. 

 In Fig.~\ref{fig_spec} we summarize the available information on 
the spectral function $\zeta(\omega)$. We plot $\zeta(\omega)/s$ 
as a function of $\omega/T$. For small $\omega$ we show the hydrodynamic
prediction in equ.~(\ref{zeta_w_h}) where $\zeta(0)$ is assumed to be 
the fluctuation bound. For large $\omega$ we show the tail predicted
by the operator product expansion, see equ.~(\ref{zeta_w_t}). We have
chosen $T/T_F=0.2$ and $\lambda/a_s=1$. We conclude that a smooth 
extrapolation of the large frequency tail to $\omega=0$ is consistent
with a bulk viscosity $\zeta(0)$ which is somewhat larger than the 
fluctuation bound. As an example we show the green dotted line which
corresponds to $\zeta=\zeta_{\it min}+\zeta_{\it micro}-c\sqrt{\omega}$
with $\zeta_{\it micro}/s=0.04$ and $c$ given by equ.~(\ref{zeta_w_t}). 
This function smoothly matches the high frequency tail. Integrating
the low frequency model and the high frequency tail over the entire
range $\omega\in [0,\infty]$ saturates 65\% of the sum rule in 
equ.~(\ref{sum_rule}). We conclude that a reasonable model of 
the bulk viscosity spectral function can be obtained by matching
the high frequency tail to the hydrodynamic spectral function combined
with a small microscopic viscosity. 

\section{Outlook}
\label{sec_out}

 In this work we have studied the role of hydrodynamic fluctuations
in the bulk stress correlation function. We have shown that fluctuations
provide a lower bound on the bulk viscosity that only depends on the 
thermal conductivity and shear viscosity as well as scale breaking in the 
equation of state. The physical mechanism for the bound can be understood
in terms of the rate of equilibration of thermal fluctuations. Consider
a fluid in equilibrium at density $\rho$ and temperature $T$. Fluctuations 
in this fluid are controlled by the entropy functional in equ.~(\ref{s_fct}).
If the fluid is compressed then the equilibrium density and temperature 
change, and as result the mean square fluctuations in $\rho,T,{\bf v}$
have to change as well. However, the mechanism for fluctuations to 
adjust involves diffusion of heat and momentum, and does not take
place instantaneously. As a consequence the fluid is slightly out 
of equilibrium, entropy increases, and the effective bulk viscosity 
is not zero. This mechanism is particularly relevant in fluids which 
have no significant microscopic sources of bulk viscosity. 

 An example of a very good fluid that does not have a simple microscopic
mechanism for generating bulk viscosity is the dilute Fermi gas near 
unitarity. Our estimates indicate that the ratio of bulk viscosity to 
entropy density near the phase transition and for $|\lambda/a_s|\gsim 1$ is 
$\zeta/s\gsim 0.1$. This is within reach of experiments involving hydrodynamic 
expansion \cite{Elliott:2013}. The effects might be even more significant 
in two dimensional gases. In these systems bulk viscosity has been studied 
using the damping of monopole oscillations \cite{Vogt:2011,Chafin:2013zca}.
It may also be possible to observe the non-analyticity of the spectral
function or the long time tail in the Kubo integrand using numerical
simulations \cite{Wlazlowski:2013owa}. 

 Our work can be extended in several directions. One interesting question
is the role of critical fluctuations in the vicinity of a second order 
phase transition \cite{Kadanoff:1968,Onuki:2002}. In that case loop 
diagrams similar to the graphs studied in this work lead to an 
enhancement of the bulk viscosity near the critical point. Another 
important problem is the study of fluctuations in relativistic fluids, see
\cite{Kovtun:2003vj,Kovtun:2011np,PeraltaRamos:2011es,Akamatsu:2016llw}. In 
that case it has been conjectured that the quark gluon plasma phase transition 
has a critical end point which is in the universality class of model H 
\cite{Hohenberg:1977ym,Son:2004iv}, and that critical fluctuations can be 
observed in the relativistic heavy ion collisions \cite{Stephanov:1998dy}.

 Acknowledgments:  This work was supported in part by the US Department 
of Energy grant DE-FG02-03ER41260 and by the BEST (Beam Energy Scan
Theory) DOE Topical Collaboration. This work was completed while 
T.~S. was a visitor at the Aspen Center for Physics, which is supported 
by National Science Foundation grant PHY-1607611.

\appendix
\section{Thermodynamic quantities}
\label{sec_app}

 We assume that the equation of state is given in the form $P=P(\mu,T)$.
A specific example is the virial expansion which provides the equation
of state in the form 
\be 
 P = \frac{\nu T}{\lambda^3} \left( z+ b_2(T)z^2 + \ldots \right)\, , 
\ee
where $\nu$ is the number of degrees of freedom ($\nu=2$ in the unitary 
Fermi gas), $\lambda=[(2\pi)/(mT)]^{1/2}$ is the thermal wave length, 
and $z=\exp(\mu/T)$ is the fugacity. Note that we have set $\hbar=
k_B=1$. Near unitarity $b_2=b_2^0+\delta b_2$ where $b_2^0=-1/(4\sqrt{2})$ 
is due to quantum statistics and \cite{Schaefer:2013oba}
\be 
\delta b_2 = \frac{1}{\sqrt{2}}\left( 
  1 + \frac{2}{\sqrt{\pi mT}a_s} + \ldots \right) \, . 
\ee
Derivatives of the pressure with respect to $\mu$ and $T$ determine the 
entropy density and pressure
\be
\label{s_and_n}
s=\left.\frac{\partial P}{\partial T}\right|_{\mu}\, , \hspace{1cm}
n=\left.\frac{\partial P}{\partial \mu}\right|_{T}\, . 
\ee
The energy density is determined by the relation
\be
{\cal E}=\mu n +sT -P\, , 
\ee
and the enthalpy per particle is $h=({\cal E}+P)/n$. In order to compute 
the specific heat at constant volume we use $V=N/n$ and write 
\bea
\label{c_v_part}
 c_{V} &=& \frac{T}{V}\left.\frac{\partial S}{\partial T}\right|_{V}
       = \frac{\partial(s,V)}{\partial(T,V)}
       = \frac{\partial(s,V)/\partial(T,\mu)}{\partial(T,V)/\partial(T,\mu)}
 \nonumber \\ 
      &=& T\left[\left.\frac{\partial s}{\partial T}\right|_{\mu}
          -\frac{[(\partial n/\partial T)|_{\mu}]^2}
                {(\partial n/\partial \mu)|_{T}}\right]\, ,
\eea
where we have defined the Jacobian 
\be 
\frac{\partial(u,v)}{\partial(x,y)} = \left|
\begin{array}{ll}
 \frac{\partial u}{\partial x} &  \frac{\partial u}{\partial y} \\
 \frac{\partial v}{\partial x} &  \frac{\partial v}{\partial y} 
\end{array}\right|\, . 
\ee
In order to compute $c_P$ we make use of the relation between 
$c_P-c_V$ and the thermal expansion coefficient $\alpha=(1/V)(\partial V/
\partial T)|_P$. This relation is given by 
\be
\label{cP}
   c_{P}-c_{V}=-\frac{T}{V}\frac{[(\partial V/\partial T)|_{P}]^{2}}
                      {(\partial V/\partial P)|_{T}}\, . 
\ee
The partial derivatives are 
\bea
  \left. \frac{1}{V}\frac{\partial V}{\partial T}\right|_{P}
   &=& \frac{1}{n}\left[\frac{s}{n}
             \left.\frac{\partial n}{\partial \mu} \right|_{T}
      -\left.\frac{\partial n}{\partial T}\right|_{\mu}\right] \, , 
  \nonumber \\
   \left. \frac{1}{V}\frac{\partial V}{\partial P}\right|_{T}
   &=& -\left. \frac{1}{n^2}\frac{\partial n}{\partial \mu}\right|_{T}\, . 
\eea
The second of these relations defines the bulk modulus $\kappa_T^{-1} = 
-V^{-1}(\partial V)/(\partial P)|_T$. We get 
\be
    c_{P}=c_{V}+T\frac{\Big[
    \frac{s}{n} (\partial n/\partial \mu)\big|_{T}
     -(\partial n/\partial T)\big|_{\mu}\Big]^{2}}
              {(\partial n/\partial \mu)\big|_{T}}\, . 
\ee
The isothermal and the adiabatic speed of sound are defined by 
\be 
 c_T^2 = \left.\frac{\partial P}{\partial \rho}\right|_T,
\hspace{1cm}
 c_s^2 = \left.\frac{\partial P}{\partial \rho}\right|_{s/n}.
\ee
We have 
\be
 c_T^2 = \frac{n}{m} \left[\left.\frac{\partial n}{\partial \mu}
\right|_T\right]^{-1}\, ,
\hspace{0.3cm}
 c_s^2 = \frac{c_P}{c_V}c_T^2\, ,
\ee
and the thermal expansion coefficient can be written as
\be 
\alpha = \frac{1}{T}\left[\frac{1}{c_T^2}\frac{T}{m} 
   \frac{c_P-c_V}{n}\right]^{1/2}\, .
\ee
Finally, we can determine the first order derivatives that appear 
in the expansion in equ.~(\ref{o_exp}). We get
\be
\left.\frac{\partial P}{\partial \rho} \right|_T
   = c_T^2 \, , \hspace{0.5cm}
\left.\frac{\partial {\cal E}}{\partial \rho} \right|_T
   = \frac{h}{m}-\frac{\alpha\kappa_T T}{\rho^2} \, , \hspace{0.5cm}
\left.\frac{\partial P}{\partial T} \right|_\rho
   = \alpha\kappa_T \, , \hspace{0.5cm}
\left.\frac{\partial {\cal E}}{\partial T} \right|_\rho
   = c_V \, ,
\ee
where $h=({\cal E}+P)/n$ is the enthalpy per particle. Partial derivatives 
of these results with respect to $T$ and $\rho$ determine the second order 
coefficients in equ.~(\ref{a_RR}-\ref{a_TT}).



\begin{thebibliography}{20}

\bibitem{Kadanoff:1963}
L.~Kadanoff, P.~Martin, 
``Hydrodynamic Equations and Correlation Function,''
Ann.\~Phys.~{\bf 24} 419 (1963).

\bibitem{Dorfman:1965}
J.~Dorfman and E.~Cohen, 
``On the density expansion of the pair distribution function for a dense 
gas not in equilibrium,''
Phys.\ Lett.\ {\bf 16} (1965) 124.

\bibitem{Ernst:1971}
M.~H.~Ernst, E.~H.~Hauge, J.~M.~J.~van Leeuwen 
``Asymptotic Time Behavior of Correlation Functions. I. Kinetic Terms,''
Phys.\ Rev.\ A {\bf 4}, 2055 (1971).

\bibitem{Pomeau:1975}
Y.~Pomeau, P.~R\'esibois, 
``Time dependent correlation functions and mode-mode coupling theories,''
Phys.\ Rep.\ {\bf 19} 63 (1975).

\bibitem{Onuki:2002}
A.~Onuki, 
``Phase Transition Dynamics,''
Cambridge University Press (2002).

\bibitem{Kovtun:2003vj} 
P.~Kovtun and L.~G.~Yaffe,
``Hydrodynamic fluctuations, long time tails, and supersymmetry,''
Phys.\ Rev.\ D {\bf 68}, 025007 (2003)
[hep-th/0303010].

\bibitem{Alder:1967}
B.~J.~Alder and T.~E.~Wainwright, 
``Velocity autocorrelations for hard spheres,''
Phys.\ Rev.\ Lett.\ {\bf 18}, 988 (1967).

\bibitem{Kadanoff:1989}
L.~P.~Kadanoff, G.~R.~McNamara, and G.~Zanetti, 
``From automata to fluid flow: Comparisons of simulation and theory,''
Phys.\ Rev.\ A {\bf 40}, 4527 (1989). 

\bibitem{Kadanoff:1968}
L.~Kadanoff and J.~Swift, 
``Transport Coefficients near the Liquid-Gas Critical Point,''
Phys.\ Rev.\ {\bf 166} (1968) 89.

\bibitem{Kovtun:2011np} 
P.~Kovtun, G.~D.~Moore and P.~Romatschke,
``The stickiness of sound: An absolute lower limit on viscosity and 
the breakdown of second order relativistic hydrodynamics,''
Phys.\ Rev.\ D {\bf 84}, 025006 (2011)
[arXiv:1104.1586 [hep-ph]].

\bibitem{Chafin:2012eq} 
C.~Chafin and T.~Sch\"afer,
``Hydrodynamic fluctuations and the minimum shear viscosity of the dilute 
Fermi gas at unitarity,''
Phys.\ Rev.\ A {\bf 87}, no. 2, 023629 (2013)
[arXiv:1209.1006 [cond-mat.quant-gas]].

\bibitem{Romatschke:2012sf} 
P.~Romatschke and R.~E.~Young,
``Implications of hydrodynamic fluctuations for the minimum shear viscosity 
of the dilute Fermi gas at unitarity,''
Phys.\ Rev.\ A {\bf 87}, no. 5, 053606 (2013)
[arXiv:1209.1604 [cond-mat.quant-gas]].

\bibitem{Landau:smII}
L.~D.~Landau, E.~M.~Lifshitz,
``Statistical Mechanics, Part II'', 
Course of Theoretical Physics, Vol.IX, 
Pergamon Press (1981).

\bibitem{Son:2005rv} 
D.~T.~Son and M.~Wingate,
``General coordinate invariance and conformal invariance in nonrelativistic 
physics: Unitary Fermi gas,''
Annals Phys.\  {\bf 321}, 197 (2006)
[cond-mat/0509786].

\bibitem{Son:2005tj}
D.~T.~Son,
``Vanishing bulk viscosities and conformal invariance of unitary Fermi gas,''
Phys.\ Rev.\ Lett.\  {\bf 98}, 020604 (2007)
[arXiv:cond-mat/0511721].

\bibitem{Chao:2010tk}
J.~Chao, M.~Braby, T.~Sch\"afer,
``Viscosity spectral functions of the dilute Fermi gas in kinetic theory,'' 
New J.\ Phys.\  {\bf 13}, 035014 (2011)
[arXiv:1012.0219 [cond-mat.quant-gas]].

\bibitem{Chao:2011cy}
J.~Chao, T.~Sch\"afer,
``Conformal symmetry and non-relativistic second order fluid dynamics,''  
Annals Phys.\  {\bf 327}, 1852 (2012)
[arXiv:1108.4979 [hep-th]].

\bibitem{Czajka:2017bod} 
A.~Czajka and S.~Jeon,
``Kubo formulae for the shear and bulk viscosity relaxation times and 
the scalar field theory shear $\tau_\pi$ calculation,''
Phys.\ Rev.\ C {\bf 95}, no. 6, 064906 (2017)
[arXiv:1701.07580 [nucl-th]].

\bibitem{Jeon:1994if} 
S.~Jeon,
``Hydrodynamic transport coefficients in relativistic scalar field theory,''
Phys.\ Rev.\ D {\bf 52}, 3591 (1995)
[hep-ph/9409250].

\bibitem{Moore:2008ws} 
G.~D.~Moore and O.~Saremi,
``Bulk viscosity and spectral functions in QCD,''
JHEP {\bf 0809}, 015 (2008)
[arXiv:0805.4201 [hep-ph]].

\bibitem{Schaefer:2013oba} 
K.~Dusling and T.~Sch\"afer,
``Bulk viscosity and conformal symmetry breaking in the dilute Fermi gas 
near unitarity,''
Phys.\ Rev.\ Lett.\  {\bf 111}, no. 12, 120603 (2013)
[arXiv:1305.4688 [cond-mat.quant-gas]].

\bibitem{Landau:smI}
L.~D.~Landau, E.~M.~Lifshitz,
``Statistical Mechanics, Part I'', 
Course of Theoretical Physics, Vol.V, 
Pergamon Press (1980).

\bibitem{Rubi:2000}
J.~M.~Rubi , P.~Mazur,
``Nonequilibrium thermodynamics and hydrodynamic fluctuations,''
Physica A {\bf 276} (2000) 477.

\bibitem{Martin:1973zz} 
P.~C.~Martin, E.~D.~Siggia and H.~A.~Rose,
``Statistical Dynamics of Classical Systems,''
Phys.\ Rev.\ A {\bf 8}, 423 (1973).

\bibitem{Ma:1976}
S.-K.~Ma,
``Modern Theory Of Critical Phenomena,''
W.~A.~Benjamin (1976).

\bibitem{Hohenberg:1977ym} 
P.~C.~Hohenberg and B.~I.~Halperin,
``Theory of Dynamic Critical Phenomena,''
Rev.\ Mod.\ Phys.\  {\bf 49}, 435 (1977).

\bibitem{DeDominicis:1977fw} 
C.~De Dominicis and L.~Peliti,
``Field Theory Renormalization and Critical Dynamics Above $T_c$: 
Helium, Antiferromagnets and Liquid Gas Systems,''
Phys.\ Rev.\ B {\bf 18}, 353 (1978).

\bibitem{Khalatnikov:1983ak} 
I.~M.~Khalatnikov, V.~V.~Lebedev and A.~I.~Sukhorukov,
``Diagram Technique For Calculating Long Wave Fluctuation Effects,''
Phys.\ Lett.\ A {\bf 94}, 271 (1983).

\bibitem{Kovtun:2012rj} 
P.~Kovtun,
``Lectures on hydrodynamic fluctuations in relativistic theories,''
J.\ Phys.\ A {\bf 45} 473001 (2012)
[arXiv:1205.5040 [hep-th]].

\bibitem{Braby:2010ec} 
M.~Braby, J.~Chao and T.~Sch\"afer,
``Thermal Conductivity and Sound Attenuation in Dilute Atomic Fermi Gases,''
Phys.\ Rev.\ A {\bf 82}, 033619 (2010)
[arXiv:1003.2601 [cond-mat.quant-gas]].

\bibitem{Hofmann:2011qs}
J.~Hofmann,
``Current response, structure factor and hydrodynamic quantities 
of a two- and three-dimensional Fermi gas from the operator product 
expansion,''
Phys.\ Rev.\ A {\bf 84}, 043603 (2011)
[arXiv:1106.6035 [cond-mat.quant-gas]].

\bibitem{Tan:2005}
S.~Tan, 
``Large momentum part of fermions with large scattering length,''
Ann.\ Phys.\ {\bf 323}, 2971 (2008)
[arXiv:cond-mat/0508320]

\bibitem{Tan:2008}
S.~Tan,
``Generalized Virial Theorem and Pressure Relation for a strongly 
correlated Fermi gas,''
Ann.\ Phys.\ {\bf 323}, 2987 (2008)
[arXiv:0803.0841].

\bibitem{Yu:2009}
Z.~Yu, G.~M.~Bruun, G.~Baym,
``Short-range correlations and entropy in ultracold atomic Fermi gases,''
Phys.\ Rev.\ A {\bf 80}, 023615 (2009)
[arXiv:0905.1836].

\bibitem{Taylor:2010ju}
E.~Taylor and M.~Randeria,
``Viscosity of strongly interacting quantum fluids: spectral functions and
sum rules,''
Phys.\ Rev.\  {\bf A81}, 053610 (2010).
[arXiv:1002.0869 [cond-mat.quant-gas]].

\bibitem{Enss:2010qh}
T.~Enss, R.~Haussmann, W.~Zwerger,
``Viscosity and scale invariance in the unitary Fermi gas,''
Annals Phys.\  {\bf 326}, 770-796 (2011).
[arXiv:1008.0007 [cond-mat.quant-gas]].

\bibitem{Goldberger:2011hh}
W.~D.~Goldberger, Z.~U.~Khandker,
``Viscosity Sum Rules at Large Scattering Lengths,''
Phys.\ Rev.\ A {\bf 85}, 013624 (2012)
[arXiv:1107.1472 [cond-mat.stat-mech]].

\bibitem{Schafer:2009dj} 
T.~Sch\"afer and D.~Teaney,
``Nearly Perfect Fluidity: From Cold Atomic Gases to Hot Quark Gluon Plasmas,''
Rept.\ Prog.\ Phys.\  {\bf 72}, 126001 (2009)
[arXiv:0904.3107 [hep-ph]].

\bibitem{Bruun:2007b}
G.~M.~Bruun, H.~Smith, 
``Shear viscosity and damping for a Fermi gas in the unitarity limit,''
Phys.\ Rev.\ A {\bf 75}, 043612 (2007)
[arXiv:cond-mat/0612460].

\bibitem{Bluhm:2017rnf} 
M.~Bluhm, J.~Hou and T.~Sch\"afer,
``Determination of the density and temperature dependence of the shear 
viscosity of a unitary Fermi gas based on hydrodynamic flow,''
arXiv:1704.03720 [cond-mat.quant-gas].

\bibitem{Ku:2011}
M.~J.~H.~Ku, A.~T.~Sommer, L.~W.~Cheuk, and M.~W.~Zwierlein,
``Revealing the Superfluid Lambda Transition in the Universal
Thermodynamics of a Unitary Fermi Gas,''
Science 335, 563 (2012)
[arXiv:1110.3309 [cond-mat.quant-gas]].

\bibitem{Elliott:2013}
E.~Elliott, J.~A.~Joseph, J.~E.~Thomas,
``Observation of conformal symmetry breaking and scale invariance in 
expanding Fermi gases,''
Phys.\ Rev.\ Lett.\ {\bf 112}, 040405 (2014)
[arXiv:1308.3162 [cond-mat.quant-gas]].

\bibitem{Vogt:2011}
E.~Vogt, M.~Feld, B.~Fr\"ohlich, D.~Pertot, M.~Koschorreck, M.~K\"ohl,
``Scale invariance and viscosity of a two-dimensional Fermi gas,''
Phys.\ Rev.\ Lett.\ {\bf 108}, 070404 (2012)
arXiv:1111.1173 [cond-mat.quant-gas].

\bibitem{Chafin:2013zca} 
C.~Chafin and T.~Sch\"afer,
``Scale breaking and fluid dynamics in a dilute two-dimensional Fermi gas,''
Phys.\ Rev.\ A {\bf 88}, 043636 (2013)
[arXiv:1308.2004 [cond-mat.quant-gas]].

\bibitem{Wlazlowski:2013owa} 
G.~Wlazlowski, P.~Magierski, A.~Bulgac and K.~J.~Roche,
``The temperature evolution of the shear viscosity in a unitary Fermi gas,''
Phys.\ Rev.\ A {\bf 88}, 013639 (2013)
[arXiv:1304.2283 [cond-mat.quant-gas]].

\bibitem{PeraltaRamos:2011es} 
J.~Peralta-Ramos and E.~Calzetta,
``Shear viscosity from thermal fluctuations in relativistic conformal 
fluid dynamics,''
JHEP {\bf 1202}, 085 (2012)
[arXiv:1109.3833 [hep-ph]].

\bibitem{Akamatsu:2016llw} 
Y.~Akamatsu, A.~Mazeliauskas and D.~Teaney,
``A kinetic regime of hydrodynamic fluctuations and long time tails for 
a Bjorken expansion,''
Phys.\ Rev.\ C {\bf 95}, no. 1, 014909 (2017)
[arXiv:1606.07742 [nucl-th]].

\bibitem{Son:2004iv} 
D.~T.~Son and M.~A.~Stephanov,
``Dynamic universality class of the QCD critical point,''
Phys.\ Rev.\ D {\bf 70}, 056001 (2004)
[hep-ph/0401052].

\bibitem{Stephanov:1998dy} 
M.~A.~Stephanov, K.~Rajagopal and E.~V.~Shuryak,
``Signatures of the tricritical point in QCD,''
Phys.\ Rev.\ Lett.\  {\bf 81}, 4816 (1998)
[hep-ph/9806219].


\end{thebibliography}
\end{document}